\documentstyle[11pt]{article}
\def\be{\begin{eqnarray}}
\def\ee{\end{eqnarray}}
\def\ben{\begin{equation}}
\def\een{\end{equation}}
\def\bi{\bibitem}

\def\calF{{\cal F}}
\def\calO{{\cal O}}\def\calR{{\cal R}}

\def\prl{Phys. Rev. Lett.}\def\np{Nucl. Phys.}
\def\pl{Phys. Lett.}
\def\J#1#2#3#4{ {#1} {\bf #2} {#3} {(#4)} }
\def\PRL{Phys. Rev. Lett.}
\def\PLB{Phys. Lett. B}
\def\NPB{Nucl. Phys. B}
\def\PRC{Phys. Rev. C}

\def\del{\partial}

\def\roughly#1{\mathrel{\raise.3ex\hbox{$#1$\kern-.75em%
\lower1ex\hbox{$\sim$}}}}\def\lsim{\roughly<}

\def\He#1{{}^{#1}\mbox{He}}\def\B#1{{}^{#1}\mbox{B}}

\def\calR{{\cal R}}

\renewcommand{\thefootnote}{\fnsymbol{footnote}}
\begin{document}
\begin{center}

  {\Large \bf Effective Field Theory For Nuclei:
\\ Confronting Fundamental Questions in
Astrophysics}\footnote{Invited talk given by MR at the
International Conference on Few-Body Problems, Taipei, Taiwan,
6-10 March 2000} \vskip 0.3cm {{\large Tae-Sun Park$^{a}$,
Kuniharu Kubodera$^{b}$,
Dong-Pil Min$^c$}\\ {\large and Mannque Rho$^{d,e}$} } 

{\it (a) Theory Group, TRIUMF, Vancouver, B.C., Canada V6T 2A3}

{\it (b) Department of Physics and Astronomy, University of South
Carolina,}

{\it Columbia, SC 29208, USA}

{\it (c) Department of Physics, Seoul National University, Seoul
151-742, Korea}

{\it (d) Service de Physique Th\'eorique, CE Saclay, 91191
Gif-sur-Yvette, France}

{\it (e) Korea Institute for Advanced Study, Seoul 133-791, Korea}

\end{center}

\vskip 0.2cm

\centerline{\bf Abstract} \vskip 0.1cm

Fundamental issues involving nuclei in the celebrated solar
neutrino problem are discussed in terms of an effective field
theory adapted to nuclear few-body systems, with a focus on the
proton fusion process and the {\it hep} process. Our strategy in
addressing these questions is to combine chiral perturbation
theory -- an effective field theory of QCD -- with an {\it
accurate} nuclear physics approach to arrive at a {\it more
effective} effective field theory that reveals {\it and} exploits
a subtle role of the chiral-symmetry scale in short-distance
effects encoded in short-range nuclear correlations. Our key
argument is drawn from the close analogy of the principal weak
matrix element figuring in the {\it hep} process to the suppressed
matrix elements in the polarized neutron-proton capture at
threshold currently being measured in the laboratories.


\renewcommand{\thefootnote}{\arabic{footnote}}
\setcounter{footnote}{0}

\section{The Challenge}
\indent\indent
 One of the currently exciting issues in astrophysics is the solar
neutrino problem, which has recently been further highlighted by
the Super-Kamiokande experiment~\cite{bah1}. 
Among the issues that have emerged
from the recent measurement is the {\it hep} process $^3{\rm
He}(p,e^+\nu_e)^4{\rm He}$.
This reaction produces highest-energy solar neutrinos
that may affect interpretations of the Super-K data
in the astrophysical context and/or in search of evidence
for new physics.
The utmost importance of the {\it hep} gives nuclear physics
a great challenge of providing reliable estimates of
the {\it hep} cross sections.
We would like to discuss this
problem in this talk in light of the recent development in
effective chiral field theories for nuclei that we (``PKMR") have
been developing for some time. Let us first state the problem
and then develop the arguments that purport to meet this
challenge.
\subsection{The {\it hep} problem}
\indent\indent
 The $hep$ process,
\be
p+\He3 \rightarrow \He4 + e^+ + \nu_e \label{hep}
\ee
 in the Sun produces neutrinos of the maximum energy
$E_\nu^{\rm max}(hep)=18.795$ MeV, which is even higher than the
maximum energy of the $\B8$ neutrino, $E_\nu^{\rm
max}(\B8)=17.980$ MeV. So the $hep$ neutrinos near the upper end
of the spectrum represent highest-energy solar neutrinos. However,
the flux of the $hep$ neutrinos is very small, because the Sun
rarely uses this weak-interaction process to produce $\He4$; the
strong-interaction process, $\He3+\He3\rightarrow\He4+2p$, can
produce $\He4$ much more efficiently. In fact, the $hep$ neutrino
flux $\phi(hep)$ is even weaker than the $\B8$ neutrino flux
$\phi(\B8)$, which is already much smaller than the primary
$pp$-$pep$ neutrino flux (see e.g., Fig. 2 of ref.\cite{bu88}). On
the other hand, since the $hep$ reaction and $\B8$ production
occur at different stages of solar burning, the detection of the
$hep$ neutrinos is expected to provide information that is not
obtainable from the $\B8$ neutrinos. This provides a strong
motivation for experimental efforts to measure the solar $hep$
neutrinos. It is hoped that the fact $ E_\nu^{\rm
max}(hep)>E_\nu^{\rm max}(\B8)$ will allow identification of the
$hep$ neutrino despite its feeble flux which is almost drowned in
the huge {\it background} of the $\B8$ neutrinos.

This spectrum has been recently measured by the Super-Kamiokande
collaboration (by measuring the energy spectrum of electrons
recoiling from scattering with the solar neutrinos)~\cite{SK}. The
analysis of the electron energy spectrum~\cite{bk98} indicates
that {\it if the observed spectrum were interpreted to be entirely
due to the {\it hep} process}, then the fit would require --
independently of neutrino oscillation scenarios -- an enhancement
in the {\it hep} $S$ factor by more than a factor of 20 relative to
the Standard-Solar-Model (SSM) value $S_0$. This would imply
that either the observed neutrinos are coming from some other
sources than the {\it hep} or an important physics ingredient is
missing that is signaling a new physics or else something is amiss
in the strong interactions.

In this talk we shall address the last possibility, namely, the
nuclear aspect of the problem. We shall thereby suggest how a
reliable bound on the $S$ factor can be given within the domain of
strong interaction physics. The SSM incorporates the presently
available ``best" matrix element of the weak current~\cite{swpc},
so the problem at hand boils down to asking whether the
discrepancy is due to our inability to calculate the nuclear
matrix element within a factor of as much as five.

Off-hand it would seem incredible that the highly successful
standard nuclear physics approach could have been so wrong and for
so long. The question is: Now that we know what the correct theory
of strong interactions that must govern nuclear dynamics involved
in the process is (i.e., QCD), why can't nuclear theorists
calculate reliably this matrix element and eliminate this big
discrepancy? The frustration associated with this question is
reflected in the recent remark by Bahcall~\cite{bah1}: ``I do not
see anyway at present to determine from experiment or from first
principles theoretical calculations a relevant, robust upper limit
to the {\it hep} production cross section (and therefore the {\it
hep} solar neutrino flux)." This then makes ``the range of values
allowed by fundamental physics for the {\it hep} production cross
section the most important unsolved problems in theoretical
nuclear physics related to the solar neutrinos."

The aim of this talk is to discuss how the above challenge could
be met by means of exploiting effective field theories of QCD
formulated for nuclear systems.
\subsection{What makes this problem so tough: Chiral filter
mechanism}
 \indent\indent
As has been forcefully argued in this conference by
Pandharipande~\cite{vijay} and Schiavilla~\cite{rocco}, properties
of light nuclei for mass number $A\lsim 9$ can be remarkably
accurately calculated in a standard (highly sophisticated)
potential-model approach with the potentials tuned to the wealth
of available data. This approach, which we shall refer to as
the ``standard nuclear physics" approach,
will later be
incorporated in a scheme that is consistent with low-energy QCD.
In measured electro-weak response functions -- with the glaring
exceptions that will be discussed below, this approach has
proven to be stunningly accurate, so the question is what makes it
so difficult to pin down the {\it hep} process matrix element
better than a factor of four or more.

One way to see what goes hay-wire in the {\it hep} process is to
use an old argument based on what is called ``chiral filter
conjecture"~\cite{KDR} which can be summarized as follows.
Whenever leading single-particle processes receive contributions
from many-body correction terms that involve one {\it soft-pion}
exchange allowed by symmetry {\it and} unsuppressed by kinematics,
then such terms -- which are calculable accurately by means of
low-energy theorems or current algebras -- will dominate and
higher-order corrections are both highly suppressed and
systematically calculable~\footnote{The well-tested processes
belonging to this class are the unpolarized cross section for
thermal neutron capture by proton $n+p\rightarrow d
+\gamma$~\cite{PMR}, electrodisintegration of the deuteron
$e+d\rightarrow n+p+e$~\cite{annrev} (both of which are dominated
by the isovector M1 operator) and weak axial charge nuclear
transitions $A(J^\pm\rightarrow J^\mp)$~\cite{PMR-axialcharge}.}.
If on the other hand, one-soft-pion-exchange terms are absent due
to symmetry or suppressed by kinematics, then higher-order
corrections to the leading one-particle processes are not
necessarily suppressed or convergent. This means that the
calculation becomes highly model-dependent and that parameter-free
predictions are not feasible. The former case will be referred to
as ``chiral-filter protected" and the latter as ``chiral-filter
unprotected."  The power of chiral filter is then to provide a
simple rule of thumb as to which processes are easily calculable
and which are not. 
However, even when processes in question are not protected
by the chiral filter,
it is in many cases still possible
to carry out sufficiently accurate calculations.
This is the case when the leading one-body terms
have substantial contributions so that, even if
next-order corrections cannot be assessed with precision,
it does not cause a major problem.
We will encounter such a situation in the proton fusion process in the Sun
\be
p+p\rightarrow d +e^+ +\nu_e.\label{pp}
 \ee
 It turns out that this process
can be calculated within an uncertainty of order of 8\%. The
reason for this is that the leading-order single-particle term
dominates although higher-order corrections, unprotected by the
chiral filter, could be uncertain by more than 100\%. The same
holds for the triton $\beta$ decay.

The situation is drastically different in the cases of the {\it
hep} process and of the suppressed isoscalar matrix elements in
the polarized $np$ capture $\vec{n}+\vec{p}\rightarrow d+\gamma$.
For the process (\ref{hep}) which undergoes primarily through an
axial weak current, the lepton momentum transfer is small, so one
would, off-hand, think that the leading term would be the allowed
Gamow-Teller (GT) matrix element. But since the GT operator just
flips spin and isospin, the matrix element involving the initial
and final hadronic states in (\ref{hep}) is suppressed by the
orthogonality of the wave functions involving different spatial
symmetries in the main components of the wave functions:
in the Young tableaux notation of the symmetry group $S_4$, the
initial state is in [31] and the final state in [4]. 
With realistic wave functions, the matrix element is not quite 
zero but tiny~\cite{swpc,rocco}. This means that 
for a reliable calculation of the process, we need to have the 
correction terms under a quantitative and systematic control.
The argument of \cite{KDR}, however,
says that they are unprotected by the 
chiral filter and hence cannot be controlled in a straightforward
way. As we will see later, the situation is further exacerbated by 
the fact that the correction terms come with an opposite sign to 
the ``leading" matrix element, causing a serious cancellation.

So how do we go about computing this process with any confidence
at all? This is the Bahcall challenge. 
We propose that a tool
to meet this challenge is to combine effective field theory
with the standard nuclear approach as presented by
\cite{vijay,rocco} and 
to formulate a {\it hybrid approach}, which we
shall call ``{\it more effective} effective field theory (MEEFT)."
\section{Effective Field Theory for Nuclei}
 \subsection{Effective field theory (EFT) defined}
\indent\indent
 The only known way to answer low-energy nuclear physics questions
from fundamental principles is to resort to effective field theory
adapted to nuclear few-body problems. 
At low energy, the relevant degrees of freedom are 
not the confined quarks and gluons of QCD but 
non-strange 
baryons and mesons. Doing a QCD calculation in
nuclear physics translates into doing an effective field theory,
which done {\it fully} is no more and no less than QCD
proper~\cite{weinbergtheorem}. So what is an effective field
theory for nuclei which we may qualify 
as ``fundamental"?

Phrased in a nut-shell and 
somewhat loosely, it goes as
follows: Suppose that we are interested in a process whose energy
(momentum) scale is much less than, say, $\Lambda$. Consider a
generic field $\Phi$ that accounts for the physics we are
interested in. Let the degrees of freedom lodged above the scale
$\Lambda$ -- in which we {\it are not} interested -- be denoted by
$\Phi_H$ and those lodged below $\Lambda$ -- in which we {\it are}
interested -- by $\Phi_L$. The physics is in the ``partition
function" $\int [d\Phi]{\rm exp} (iS(\Phi))$. Since we are not
interested in $\Phi_H$ we may integrate it out and express the
partition function in terms of an effective action $S^{eff}
(\Phi_L)$ as $\int [d\Phi_L]{\rm exp} (iS^{eff} (\Phi_L))$ where
${\rm exp} (iS^{eff})=\int [d\Phi_H] {\rm exp}
(iS(\Phi_H,\Phi_L))$. This is just a formal manipulation, so
nothing will be gained unless one can arrive at a manageable
$S^{eff}$ -- expressed in terms of a reduced number of relevant
degrees of freedom that we can control -- with which we could do
physics systematically. The next step is to approximate the
effective action, in general {\it non-local}, by an infinite sum
of {\it local} terms as
\be
S^{eff} (\Phi_L)=\sum C_n \calO_n \ee
 where $\calO$'s are local operators involving the field $\Phi_L$
and $C$'s are coefficients dependent on the scale $\Lambda$ to be
determined either theoretically from first principles or from 
experiments. Since physical quantities should not depend upon
where one puts the separation or cutoff $\Lambda$, the
coefficients $C_n$ should satisfy the renormalization-group (RG)
flow equation (known as
the
Wilson RG equation)
 \be
 \del C_n (\Lambda)/\del\Lambda=\calF (\Lambda).\label{rge}
 \ee
 If the theory is completely defined everywhere, that is, both
above and below $\Lambda$, then one may do the integrating-out
explicitly and obtain an effective action. By a procedure of
``matching,"  one could fix $C_n$ term by term. In nuclear
physics, this is not possible since the relevant degrees of
freedom are not the same in the quark-gluon and hadron regimes and
there is no known way to go from one to the other. In this case
the best one can do is to truncate the series at low enough order
and resort to a certain number of ``clever guesses." It is at this
stage that one needs to have a guiding principle 
for making 
right guesses and reliable error estimates.
This necessarily injects a certain degree of ``art" in the actual
calculation.

If the right-hand side of eq.(\ref{rge}) goes to zero at some
$\Lambda$, then it is a signal that there is a fixed point. This
is the case for Landau Fermi liquid parameters~\cite{fermiliquid}
in many-body systems. An effective field theory for nuclear matter
that exploits the Fermi-liquid fixed points and BR
scaling~\cite{BR91} has been proposed in \cite{song}. In the
problem at hand, this aspect plays no role, so we will not discuss
it any further.
\subsection{{\it More effective} EFT}
 \indent\indent
In dealing with electro-weak response functions and more
significantly with $n$-body systems with $n>1$, we find the hybrid
approach developed over the years by PKMR ~\cite{PKMR} to be more
predictive than its cousins~\cite{KSW} and in some cases to be the
{\it only} predictive method, although for two-body systems, they
are more or less equivalent. The hybrid approach -- which follows
Weinberg's proposal~\cite{weinberg,bira} -- consists of
distinguishing two sets of graphs, one ``irreducible graphs" and
the other ``reducible graphs." In treating the irreducible graphs
that involve no infrared enhancements, one uses standard chiral
perturbation theory. This generically involves an electro-weak
vertex $\Gamma$ (which enters at most once as we will be dealing
with a slowly varying external field) and a two- or many-body
interaction potential $V$ computed to a finite order of chiral
expansion. The reducible graphs involving $V$ and
infrared-enhanced propagators on the other hand need to be
iterated to all orders, a procedure which is effectively executed
in Lippman-Schwinger equation or Schr\"odinger equation. The
hybrid strategy is then to take the accurate wave functions
generated by the procedure as described by Pandharipande and
Schiavilla in this meeting and compute the relevant matrix
elements with the current vertex computed in chiral perturbation
theory to as high an order as feasible which in practice turns out
to be next-to-next-to leading order. Beyond that order, the theory
becomes unpredictive due to the paucity of experimental data.

We should mention a caveat to this (hybrid) procedure: Since the
reducible graphs are iterated to all orders while the irreducible
graphs (i.e., current vertex $\Gamma$ and potential $V$) are
computed to a finite order $n$, one makes an error at order $n+1$
in the EFT counting. This caveat can be avoided in principle as
e.g. in the approach advocated by the authors in \cite{KSW}.
However in practice, the counting error at the $(n+1)$ order makes
a negligible numerical error in all chiral-filter-protected
observables. In fact one can show rather convincingly that in
two-body systems at low energy, the two methods are essentially
equivalent~\cite{cohen}.

We propose that this hybrid theory is {\it more effective} than an
effective theory of the type advocated in \cite{KSW} for certain
processes. The ingenuity is to realize that one can turn the above
caveat around to advantage when one is calculating three- or
more-body systems and, perhaps more significantly, when the
chiral-filter protection is simply missing as in the {\it hep} and
$np$ processes.
\subsection{Power counting}
 \indent\indent
Computing the irreducible current vertex $\Gamma$ and the
potential $V$ is systematically organized by a power counting rule
which in the present case is the usual chiral
counting~\cite{weinberg}. Let the characteristic probe
energy/momentum scale be given by $Q$ and the cutoff that
separates  the ``high (H)" and ``low (L)" regimes be $\Lambda$. We
are interested in making an expansion in $Q/\Lambda$ with a series
$C_\nu (Q/\Lambda)^\nu$. In {\it more effective} EFT (MEEFT), the
potential is taken from nature -- a procedure which is consistent
with the tenet of effective field theories contrary to
misunderstandings among some theorists , so it suffices for our
purpose to focus solely on the current vertex. The index $\nu$ for
an $n_B$-body current is given by~\cite{weinberg}
 \be
\nu=2(n_B-1)+2L+\sum_i \nu_i
 \ee
 where
 \be
 \nu_i=d_i+e_i+n_i/2-2
 \ee
and where $L$ is the number of loops, $n_i$ the number of nucleon
lines entering the $i$ th vertex, $d_i$ the number of derivatives
and $e_i$ the number of external fields. We are dealing with a
slowly varying field so the external field acts only once, i.e.,
$e_i=1$.

In discussing higher order corrections, we shall denote
by N$^\nu$LO the $\nu$-the order correction 
{\it relative} to the LO.

It is worth noting at this point that relative to one-body
currents, the leading 2-body currents are N$^1$LO if protected by
the chiral filter and N$^3$LO if not. This is because the sum
$\sum_i\nu_i$ is different by 1 between the LO 1-body and LO
2-body currents.~\footnote{For example,  if we look at
$\sum_i\nu_i$ for the (LO 1-body, LO 2-body) contributions, we
find (1, 0) for the isovector M1 and the axial charge, (1, 2) for
the isoscalar M1, and (0, 1) for the GT.}

\subsection{Chiral constraint}
\indent\indent
 Chiral symmetry has an important constraint on
$\nu_i$, namely that $\nu_i\ge 0$. The presence of an external
field in this constraint has a non-trivial consequence on the
chiral filter as pointed out in \cite{mr91}. This is because power
counting is subtle when an external field figures. In the Weinberg
counting~\cite{weinberg} which we adopt here and in the absence of
external fields (that is, in the potential), the leading one-pion
exchange and leading four-Fermi contact interaction terms are of
the same chiral order, so the long-range one-pion exchange and the
contact (short-range) interaction are equally important. However
when an external field is present, the contact interaction becomes
sub-dominant to one-pion exchange~\cite{mr91}. This means that
short-range many-body currents are of higher order in chiral
counting than a one-pion-exchange current. This obvious point
turns out to be essential for understanding the chiral filter
mechanism. A consequence is that if one were to calculate $n$-body
currents by insisting that they be strictly consistent with the
corresponding potential (e.g., by imposing current conservation),
then to be consistent {\it also} with chiral symmetry, other terms
of the same chiral order as the short-ranged current -- such as
{\it pion loops and counter terms} -- must be included. More
specifically, this implies that the procedure often used in the
literature to calculate two-body currents with the exchange of all
the mesons that figure in the one-boson-exchange potential model,
namely, $\pi$, $\rho$, $\omega$, $\sigma$ etc. is incomplete and
hence could be erroneous numerically unless supplemented with
additional counter terms and loop terms of the same chiral orders.
\subsection{Two scales}
\indent\indent
 There are basically two relevant scales in the problem: ``nuclear
scale" and ``chiral scale." In the low-energy regime we are
concerned with, the nuclear scale is given by one or two pion
masses depending upon whether one introduces the pion explicitly
or integrates it out. A viable theory must be stable around this
 nuclear scale. It has been verified that this stability is easily
 assured by all consistent EFTs~\cite{PKMR,KSW}.

 The chiral scale is essentially given by spontaneously broken
chiral symmetry characterized by the scale $\sim 4\pi m_\pi\sim
m_N\sim 1$ GeV. 
This scale delineates vector mesons $\rho$,
$\omega$, glueballs etc. from the 
low-energy regime and short-distance
physics from long-distance physics. In our treatment, nucleons and
pions are introduced explicitly while all others are integrated
out, so the chiral scale basically corresponds to the scale at
which short-range correlation of the standard nuclear physics
approach enters, namely at the scale of vector meson masses.
\section{Polarized Neutron-Proton Capture: Hint for a Strategy}
\subsection{Hard-core cutoff scheme (HCCS)}
 \indent\indent
The capture cross section for $n+p\rightarrow d+\gamma$ is
dominated by an isovector $M1$ operator (denoted $M1V$) which is
chiral-filter protected according to \cite{KDR,mr91} and hence is
very accurately predicted~\cite{PMR}. 
Meanwhile,
the matrix elements of the
isoscalar electromagentic current are 
suppressed by a factor of $\sim 10^3$. 
Thus they cannot be ``seen" in the
total cross section. However the isoscalar $M1$ (called $M1S$ to
distinguish it from the isovector $M1$) and $E2S$ 
(isoscalar E2)
can be extracted
via spin-dependent observables from the polarized capture process
 \be
 \vec{n}+\vec{p}\rightarrow d+\gamma\label{polnp}
 \ee
which is being measured at Institut Laue-Langevin (ILL) in
Grenoble~\cite{mueller}.

We will now make predictions for these matrix
elements~\cite{PKMR-ILL}. In doing so, we will learn how to
compute the {\it hep} process. There is a very close parallel
between the two.

Very much like in the {\it hep} process, the $M1S$ matrix element
is highly suppressed due to the orthogonality between the initial
wave function $^3S_1$ and the final deuteron wave function. This
makes the $M1S$ as suppressed as $E2S$.
Furthermore the isoscalar EM current is {\it not} protected by the
chiral-filter mechanism, so the corrections can be uncomfortably
large and non-controllable.

There is a way out of this impasse and it is found in the
exploitation of the two ingredients of the MEEFT~\cite{PKMR-ILL}.
Firstly the initial and final wave functions are very accurately
known~\cite{argonne} and secondly to next-to-next-to leading order
(N$^4$LO) (where the leading order (LO) corresponds to the
single-particle $M1S$ matrix element), the loops and counter terms
required by chiral symmetry are controlled by the chiral cutoff
$\Lambda$ reflecting short-distance physics involving a scale
$\Lambda \sim 4\pi f_\pi\sim m_N\sim 1$ GeV. There are four
independent counter terms that arise at N$^3$LO and N$^4$LO. Most
remarkably, {\it to the order we are working in, they combine to a
single constant $X$ in the precise form that is required to fix
completely the deuteron magnetic moment for a given $\Lambda$.} 
We stress that, were 
it not for this ``coincidence,"  there would be
too many undetermined parameters and it would be impossible to
make a prediction for $M1S$. It should also be mentioned that this
unique correspondence is probably possible only to N$^4$LO. We
suspect that to higher orders, the one-to-one correspondence that
makes the prediction possible would be spoiled.

Now the tenet of a viable EFT is that the result should not depend
on the cutoff $\Lambda$ as it represents the scale at which
$\Phi_H$ and $\Phi_L$ are separated -- which is arbitrary. The
degrees of freedom that have been integrated out in
\cite{PKMR-ILL} are heavy mesons (such as $\rho$, $\omega$,
glueballs etc.). Any strong dependence on $\Lambda$ would
therefore signal that certain short-distance physics is missed
and/or that we have a non-converging series. {\it What we have
done in fixing the counter terms at a given $\Lambda$ to the
empirical magnetic moment of the deuteron corresponds to a
specific regularization scheme called `` hard-core cutoff scheme
(HCCS)"}\footnote{In \cite{PKMR-ILL}, this scheme was referred to
as ``MHCCS."} which subsumes certain terms higher order than
N$^4$LO as required by the empirical deuteron magnetic moment. In
principle, the counter terms are {\it presumably} calculable (in
the far future) order by order from ``first principles" for a
given $\Lambda$. However since we are using a particular
regularization scheme, the counter terms we fix from experiments
are {\it not} necessarily calculable order by order.
\subsection{Prediction}
 \indent\indent
What the experimenters are measuring are the ratios $\calR$, i.e.,
the $M1S$ and $E2S$ relative to the isovector M1 ($M1V$),
\be
\calR_{M1S}=\frac{M1S}{M1V},\ \ \ \calR_{E2S}=\frac{E2S}{M1V}.
 \ee
In \cite{PKMR-ILL}, regularization was effected with the distance
cutoff $r_c$ instead of a momentum cutoff. This cutoff is closely
related to the ``hard-core radius" used in the standard nuclear
physics approach. The range of $r_c$ we have explored was
\be
r_c^{min}\equiv 0.01\ {\rm fm} \le r_c \le r_c^{max}\equiv 0.8\
{\rm fm}.
 \ee
This is wide enough to amply cover the range implied by
spontaneously broken chiral symmetry. Expressed in the chiral
order and in the range $(r_c^{max},r_c^{min})$,  the $M1S$ comes
out to be
\be
\calR_{M1S}\times 10^3 &=&``{\rm LO}"  + \ \ \ \ ``{\rm N}^3{\rm
LO}" \ \ + \ \  ``{\rm N}^4{\rm LO}" \ \  + \ ``({\rm N}^3{\rm LO}
+{\rm N}^4{\rm LO})_X"\nonumber\\
 &=& -0.74+(-0.48,-0.74)+(0.23,0.46)+(0.49,0.54)\nonumber\\
 &=& (-0.50,-0.49)
 \ee
where the contribution that depends on the single parameter $X$
fixed  by the deuteron magnetic moment is indicated by the
subscript $X$. The crucial observation here as well as later in
the case of the {\it hep} process is that the final result cannot
be arrived at by any partial sum of the terms. Note also the
important role of the $X$-dependent term.

The $E2S$ on the other hand does not depend on $X$ and the
correction from higher orders (N$^3$LO and N$^4$LO) to the LO
result is negligible
 \be
 \calR_{E2S}\times 10^{3}=0.24+ \calO (10^{-3}).
 \ee

The lessons we learn from this prediction are two-fold. One,
because of the strong suppression of the LO term, the corrections
are all comparable independently of the chiral order and not small
compared to the LO. Second, the final result is practically
independent of the cutoff $r_c$, assuring that no important part
of short-distance physics -- presumably encoded in short-range
correlation -- is missed in the procedure. It suggests that the
regularization procedure adopted here in some sense {\it forces}
the series to converge at N$^4$LO.\footnote{It is quite remarkable
that the same results were obtained in a different procedure by
Chen, Rupak and Savage~\cite{crs}, i.e., $\calR_{M1S}=-0.50 (1\pm
0.6)$ and $\calR_{E2S}=0.25 (1\pm 0.15)$.}
\section{Solar Neutrino Processes}
\subsection{Chiral filter protection/non-protection}
\indent\indent
 Since the lepton momentum transfer in the
process (\ref{hep}) -- which is mediated by the isovector axial
current $A_\mu^a$ -- is small, the leading-order (LO) term in the
current is the one-body Gamow-Teller (GT) operator with index
$\nu=0$ coming from the space component $\vec{A}^a$. The
next-order terms are first-forbidden corrections to the one-body
GT operator and the axial charge one-body contribution. In the
results reported here, the former will be incorporated into the LO
term. The latter will be treated separately.

The axial-charge operator $A_0^a$ is {\it protected} by chiral
filter~\cite{KDR}: corrections to the one-body axial charge
operator (of order $\nu=2$ and N$^1$LO) are dominated by
one-soft-pion 
exchange. Thus corrections to order $\nu=2$ are completely fixed. 
Further corrections to the axial charge are strongly suppressed 
and can be ignored~\cite{PMR-axialcharge}.

On the contrary, the space component $\vec{A}^a$ in which the
one-body GT operator is the ``leading" term is {\it not} protected
and so the $n$-body corrections are generically uncontrollable. It
is this part that challenges the theorists. Most fortunately,
however, it turns out that the procedure exploited for the
neutron-proton capture process applies in a complete parallel to
the {\it hep} process.

\subsection{The proton fusion}
\indent\indent
 Before we tackle the {\it hep} problem, consider the proton fusion
 process (\ref{pp}) which is predominantly given by
the unsuppressed one-body Gamow-Teller (GT) operator. For this,
the axial-charge operator does not figure. Since the space
component of the axial current is not protected by the chiral
filter, corrections to the GT term -- that are two-body -- can be
calculated only to N$^3$LO which involve no loops and no counter
terms. To N$^3$LO, one obtains~\cite{PKMR-astro}
\be
M_{GT}=4.77 (1+0.04).\label{gt}
 \ee
 Although the N$^3$LO correction (given in the parenthesis in
(\ref{gt})) is only 4\%, there is nothing that would suggest that
the N$^4$LO and higher order terms can be ignored compared to the
N$^3$LO.  Unfortunately to N$^4$LO, there are too many unknown
counter terms to make an unambiguous estimate. The error could
well be 100\% or more of the N$^3$LO term.

Even so, given that the one-body GT term is dominant, it still
makes sense to compute the solar $S$ factor using
(\ref{gt})~\footnote{The ``error" quoted here represents a range
of uncertainty involved at N$^4$LO which is not presently
calculable in a parameter-free way and should not be taken
quantitatively.},
\be
S_{pp}^{th}=4.05 (1\pm 0.08)\times 10^{-25} {\rm MeV b}
 \ee
which is close to the value adopted in the standard solar
model~\cite{bahcalletal}
\be
S_{pp}^{SSM}=4.00 (1\pm 0.007^{+0.020}_{-0.011})\times 10^{-25}
{\rm MeV b}.
 \ee
 This result illustrates a case where the leading term is unsuppressed
and corrections to it are sufficiently small so that the chiral
filter non-protection does not spoil 
the predictive power of the calculation
although
greater accuracy cannot be achieved 
without further experimental
inputs that are not expected to be forthcoming in the near future.
A recent calculation~\cite{kong} using the approach of \cite{KSW}
supports the prediction of \cite{PKMR-astro} although because of
one free parameter, the result of \cite{kong} is not, strictly
speaking, a genuine prediction.

We note without any details that a similar situation holds for the
triton beta decay. There are four parameters at N$^3$LO and
N$^4$LO that cannot be determined at present either by theory or
by experiments. It turns out however that they can be combined
into one parameter -- denoted $Y$ -- that enters into the {\it
hep} matrix element.
\subsection{The {\it hep} process}
\indent\indent
 As mentioned the axial-charge contribution is unambiguously
calculable, so we will not go into it: We see no obstacle to
calculating it with accuracy. We will therefore focus on the main
quantity at issue, the GT term and $n$-body corrections thereof.
A precise calculation of these quantities is possible because,
as in the case of $M1S$ in the neutron-proton capture, the
counter terms -- there are four of them -- at the N$^3$LO and
N$^4$LO order can be combined precisely into a single constant $Y$
that supplies the only undetermined constant present at N$^4$LO in
the triton beta decay $^3H\rightarrow ^3He +e^- +\nu_e$. By fixing
the constant $Y$ from the triton beta decay, the theory for the
controversial term in the {\it hep} process is completely
determined. We have verified this only up to N$^4$LO, beyond
which, we suspect, such a ``miracle" may cease to be operative.
Expressed in the ascending chiral order and in the momentum cutoff
range $(\Lambda_{min}=0.5 {\rm GeV},\Lambda_{max}=1.0 {\rm GeV})$,
we find with the Av14 wave function~\footnote{One can perhaps do
better with the Av18 wave functions as in \cite{rocco} but we
believe what we used is good enough for the present purpose.}
 \be
M_{GT}/g_A &=&``{\rm LO}"  + \ \ \ ``{\rm N}^3{\rm LO}" \ \ + \ \
``{\rm N}^4{\rm LO}" \ + \ ``({\rm N}^3{\rm LO} +{\rm N}^4{\rm
LO})_Y" +``{\rm N}^4{\rm LO}_{3body}"\nonumber\\ &=&
-0.38+(0.69,1.50)+ (0.24,-0.94)+\ (-0.74,-0.43)\ +\ \ \
(0.002,0.004)\nonumber\\
 &=& (-0.19,-0.24).
 \ee
 Note that as mentioned, the LO and N$^3$LO terms come with opposite
signs resulting in the cancellation mentioned before. Although
negligible, three-body currents that appear at N$^4$LO are
included for comparison. There are no loop contributions at this
order and since only pion exchanges are involved, no unknown
constants are involved. 
It should be noticed that as in the case
of the suppressed isoscalar
matrix elements in the $np$ capture, the
$Y$-dependent contribution is crucial.

Our final value with an error estimate is~\cite{PKMR-hep}
\be
M_{GT}/g_A=-0.20\pm 0.06.\label{hep-prediction}
 \ee

As in the $np$ case,
due to the strong suppression
of the leading term,
all contributions come in with more or less
equal strengths. Furthermore as in the case of the polarized $np$
capture, no partial sum of the terms resembles the final result.
It is only in the total that the strong $\Lambda$ dependence
disappears. 
Like in the polarized $np$ case,
we interpret the approximate $\Lambda$-independence
(for the wide range of $\Lambda$ considered) as
an indication that short-distance physics is properly
captured in the regularization scheme.
\section{Implication On The {\it hep} Problem}
\indent\indent
 In this note, we have suggested how to pin down the principal
matrix element governing the process (\ref{hep}) that has
frustrated for years the effort to calculate a reliable bound for
the {\it hep} $S$ factor. Once we obtain the chiral-filtered and
hence accurately calculable contribution from the axial charge
operator, we will be in a position to confront directly the
Super-Kamiokande data from the point of view of strong-interaction
physics.

While awaiting the numerical value of the axial-charge matrix
element as well as the re-confirmation of (\ref{hep-prediction})
(in progress at the time of this writing), we can still make a
reasonable qualitative statement. Since the axial-charge
contribution is of the same order as the first-forbidden term, it
cannot be much greater than the GT term. Tenuous though it might
be, some support for this conjecture comes from the recent result
of Marcucci et al~\cite{rocco} who estimated the axial-charge
two-body currents to contribute about 40\% to the $S$ factor. It
should however be cautioned that since their approach to the
axial-charge operators is different from ours, we cannot carry it
over directly to our result. Be that as it may, for the sake of
argument, we shall assign a generous error of 100\% to the matrix
element that we computed to account for the axial-charge
contribution and possible first-forbidden corrections that still
need to be accounted for. This would give the ratio of the
predicted $S$ factor $S^{th}$ to the Super-Kamiokande observation
$S^{SK}$
\be
\frac{S^{th}}{S^{SK}}\lsim 0.10\sim 0.12
\ee
 more or less independently of the neutrino oscillation
 scenarios~\cite{bah1}
 (i.e., MSW or vacuum). There is still an order of magnitude
 ``missing" in the $S$ factor which cannot be accounted for by
 strong interaction mechanisms only.

 The conclusion then is either (1) the Super-Kamiokande experiment
is wrong, (2) the experiment is correct but the interpretation of
the data is incorrect (e.g., the observed events could be due to
``backgrounds" such as supernovae) or (3) there is new physics in
the sectors outside of strong interactions. Which of these
alternatives is the viable one could probably be settled by such
laboratory experiments~\cite{ABG} as $n+^3{\rm He}\rightarrow
^4{\rm He}+e^-+\bar{\nu}_e$ and $e^- +^4{\rm He}\rightarrow ^3{\rm
H} +n+\nu_e$.

\end{document}